\documentclass[pra,amsmath,amssymb,twocolumn]{revtex4-1}
\usepackage{graphicx}
\usepackage{dcolumn}
\usepackage{bm}
\usepackage[utf8]{inputenc}
\usepackage[english]{babel}
\usepackage{amsfonts}
\usepackage{hyperref}
\usepackage{physics}
\usepackage{epstopdf}
\usepackage{mathtools}
\usepackage[capitalise]{cleveref}
\usepackage{float}
\usepackage[space]{grffile}
\usepackage{color}
\usepackage[normalem]{ulem}
\usepackage{subfigure}

\binoppenalty=\maxdimen
\relpenalty=\maxdimen

\addto\captionsenglish{}

\def \htx{\hat{\tilde{x}}}
\def \htp{\hat{\tilde{p}}}
\def \hta{\hat{\tilde{a}}}

\def \ts{\tilde{s}}
\def \G0{\tilde{G}_0^R}

\def \hx{\hat{x}}

\def \hp{\hat{p}}

\def \ha{\hat{a}}

\def \hc{\hat{c}}

\def \hH{\hat{H}}

\newcommand{\create}[1]{#1^\dagger}          

\begin{document}
	
	\title{Phase-dependent chiral transport and effective non-Hermitian dynamics in a bosonic Kitaev-Majorana chain}
	
	\author{A. McDonald$^{1,2}$, T. Pereg-Barnea$^{1,3}$ and A. A. Clerk$^2$}
	\affiliation{$^1$Department of Physics, McGill University, Montr\'eal, Qu\'ebec, Canada.\\
		$^2$Institute for Molecular Engineering, University of Chicago, Chicago, IL, USA.\\
		$^3$Department of Condensed Matter Physics, Weizmann Institute of Science, Rehovot, 76100, Israel}
	

	\begin{abstract}
		We study a 1D chain of non-interacting bosonic cavities which are subject to nearest-neighbour parametric driving. With a suitable choice of drive phases, this model is strongly analogous to the celebrated Kitaev chain model of a 1D p-wave superconductor. The system exhibits phase-dependent chirality:  photons propagate and are amplified in a direction that is determined by the phase of the initial drive or excitation.  Further, we find a drastic sensitivity to boundary conditions: for a range of parameters, the boundary-less system has only delocalized, dynamically unstable modes, while a finite open chain is described by localized, dynamically stable modes. While our model is described by a Hermitian Hamiltonian, we show that it has a surprising connection to non-Hermitian asymmetric-hopping models.  
	\end{abstract}

	\maketitle
	
	Superconducting fermionic systems are by now well known to exhibit unique kinds of topological behaviour.  Perhaps the best known example is the Kitaev chain model \cite{Kitaev2001}, the simplest possible model of a spinless p-wave superconductor:  one takes a 1D tight-binding chain of spinless fermions with nearest neighbour hopping, and adds pairing terms on each bond.  This model exhibits topologically protected end Majorana zero modes, and underpins the current quest to realize  Majorana-based topological quantum computation \cite{Alicea2012,BeenakkerReview,DasSarma2015,Oreg,Karzig}.
	
	Parametrically-driven bosonic systems have quadratic Hamiltonians with the same basic form as that of a fermionic superconductor; recent research has shown that they too can exhibit unique topological behaviour.  The parametric drive corresponds to coherent two photon addition and removal, and is analogous to a superconducting pair potential.  The lack of any exclusion principle implies that such bosonic models can differ strongly from their fermionic counterparts. Recent work has demonstrated that non-interacting parametrically-driven models can realize unique forms of topological phases, with bands characterized by an integer Chern number, and with edge states that act as ideal amplifying/squeezing channels \cite{Barnett2013,Shindou2013,Galilo2015,Brandes2015,Bardyn2016,Peano2016,Peano2016b,Brandes2016}.  Work has also shown that in 1D, strong interactions fermionize bosons, letting parametric drives directly realize the fermionic Kitaev chain \cite{Bardyn2012}.
	
	In this work, we consider a one dimensional, parametrically driven system whose quadratic Hamiltonian has an almost identical form to the fermionic Kitaev chain.
	Unlike Ref.~\cite{Bardyn2012}, we do not require strong on-site repulsive interactions, but instead consider the effects of breaking inversion symmetry.  Remarkably, we find that this system has striking features reminiscent of the Kitaev chain:  the system is best understood in terms of Hermitian quadrature operators, not photonic annihilation operators, and exhibits a strong sensitivity to boundary conditions, including the presence of highly localized states. The system also has unique transport properties: photons propagate and are amplified in a chiral fashion, with the direction being set by the {\it phase} of the initial excitation.  As we discuss, this is reminiscent of the behaviour of edge modes in the spin Hall effect\cite{Kane-Mele-Sep,Kane-Mele-Nov,Maciejko:2011kr}.
	We also point out a surprising connection between our model and non-Hermitian asymmetric hopping models \cite{Hatano1997}.
	
	In what follows, we introduce our model, outline its exceptional properties, and discuss implementations based on superconducting quantum circuits. 
	\begin{figure}
		\centering
		\includegraphics[width=0.45\textwidth]{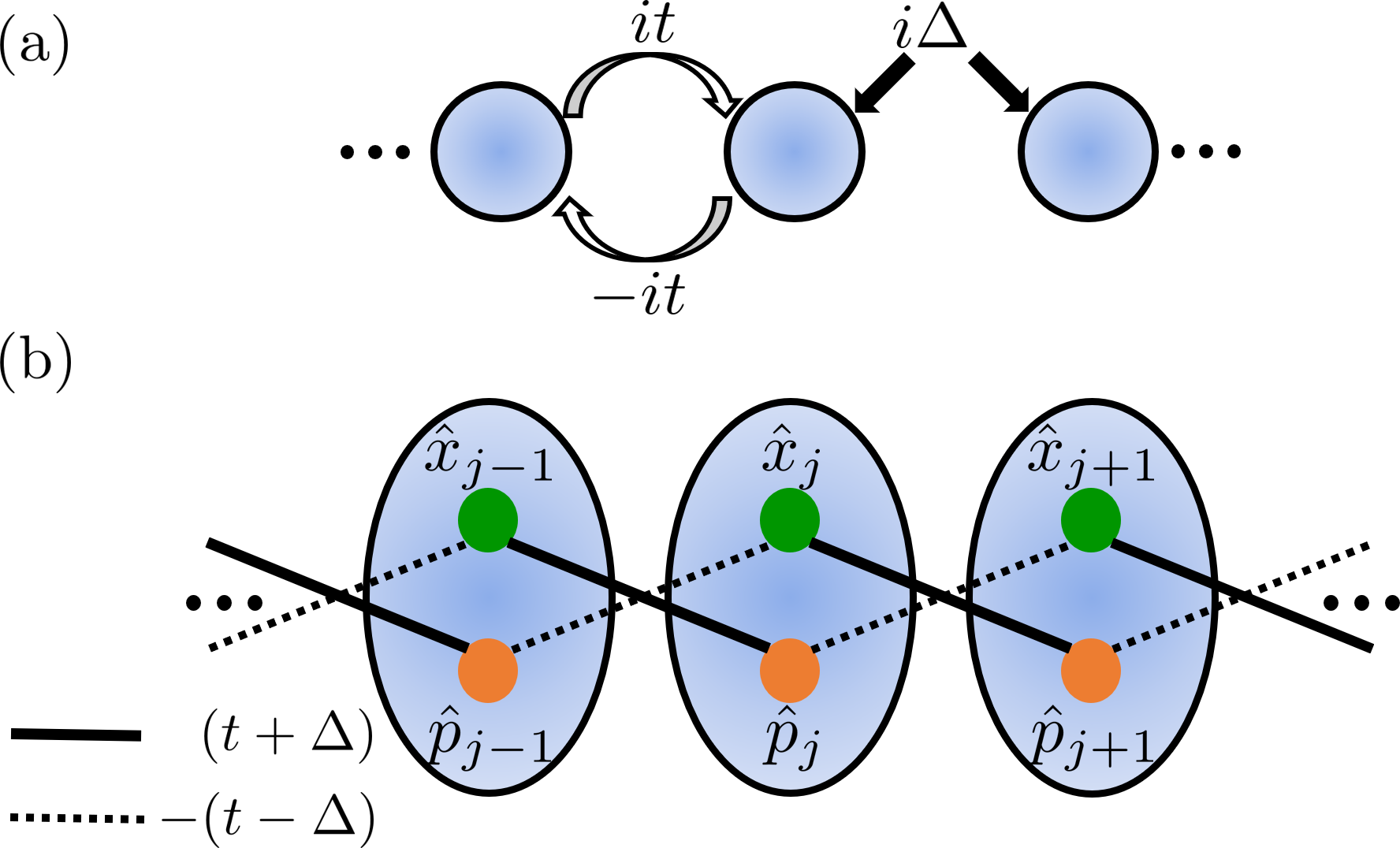}
		\caption{(a)  Schematic of the setup: an array of tunnel-coupled cavities (hopping $t$) is subject to resonant parametric driving (amplitude $\Delta$) on each bond.  We require a purely imaginary hopping matrix element, which is gauge-equivalent to having staggered parametric drive phases.  (b) When written in terms of local cavity quadratures $\hx_j$ and $\hp_j$, the model describes a spatially asymmetric pattern of couplings, strongly reminiscent of the asymmetric coupling of Majorana operators in the fermionic Kitaev-Majorana chain.} 
		\label{fig:Schematic}
	\end{figure}
	\section{Model}
	We start by recalling the fermionic Kitaev model \cite{Kitaev2001}: spinless electrons on a 1D lattice subject to a $p$-wave pairing potential.   In momentum space and at zero chemical potential it takes the form 
	\begin{equation}
	\label{eq:HFermion}
	\hH_{F} =  \sum_{k } 
	\left[
	t \cos k \,  \hc^\dagger_k \hc_k 
	+
	i \frac{\Delta}{2}  \sin k  \left(
	\hc^{\dagger}_{k} \hc_{-k}^\dagger - h.c.  
	\right)
	\right],
	\end{equation}
	where $\hc_k$ is a fermionic annihilation operator, and the sum runs over the first Brillouin zone;  we take $t>0, \Delta >0$ throughout.  The remarkable features of this model can be related to its topological properties.
	To see this, we define $\create{C}_k = (\create{c}_k, c_{-k})$ and write Eq.~(\ref{eq:HFermion}) as
	\begin{equation}
	\label{eq:HFermion_Wind}
	\hat{H}_F = \frac{1}{2}\sum_{k}\create{C}_k \left(\mathbf{h}_F(k) \cdot 				\check{\boldsymbol{\sigma}}\right)C_{k},
	\end{equation}
	where $\mathbf{h}_F(k) = (0, -\Delta \sin(k),t\cos(k))$ and $\check{\boldsymbol{\sigma}}$ is a vector of Pauli matrices in particle-hole space. The Hamiltonian $\hat{H}_F$ has chiral symmetry as the first component of the vector $\mathbf{h}_F(k)$ is zero for all momenta.  This allows us to define a topological number associated with the number of times that $\mathbf{h}_F(k)$ encircles the origin. In the Kitaev model, the topological number is non-zero as the vector $\mathbf{h}_F(k)$ winds once.
	
	Can we construct a similar bosonic model with momentum space winding?  At first glance, the answer is no. Simply replacing fermionic operators in Eq.~(\ref{eq:HFermion}) with bosonic ones does not work:  since bosonic operators commute (not anti-commute), the two-photon term must now be an even function of $k$, and thus cannot be proportional to $\sin k$.  As a result, a bosonic model with the identical real-space form as the  fermionic Kitaev chain has no non-trivial topology, and no unusual properties (see Appendix \ref{app:Trivial}).
	
	To obtain a bosonic model that has a non-zero winding in momentum space, we can keep the pairing term even in $k$, but instead make the kinetic energy vary as $\sin k$:
	\begin{equation}
	\label{eq:Hk}
	\hH_{B} =  \sum_{k } 
	\left[
	t \sin k  \,
	\ha^\dagger_k \ha_k 
	+
	i \frac{\Delta}{2} \cos k  \left(
	\ha^{\dagger}_{k} \ha_{-k}^\dagger - h.c.  
	\right)
	\right], 
	\end{equation}
	where $\mathbf{h}_F(k)$ from the fermionic case is replaced by $\mathbf{h}_B(k) = \left(0,-{\Delta} \cos k,t\sin k\right)$. The form of $\hH_B$ in real space,
	\begin{equation}
	\label{eq:Hreal}
	\hH_B = \frac{1}{2} \sum_{j }
	\left( i t  \ha^{\dagger}_{j+1} \ha_j + 
	i \Delta  \ha^{\dagger}_{j+1} \ha_{j}^\dagger + h.c. \right),
	\end{equation}
	is almost identical to that of the Kitaev chain, except that we now have a purely imaginary hopping matrix element.
	Due to the non-zero pair potential, this $\pi/2$ phase cannot be gauged away. It breaks the symmetry of the Hamiltonian under spatial inversion (i.e.~the operation taking site $j$ to $-j$). 
	As we will now explore in detail, this model has a number of remarkable properties.  
	The case of an arbitrary phase for the hopping matrix element is treated in Appendix \ref{app:ArbPhase}; for a range of parameters such models exhibit analogous physics to the model in Eq.~(\ref{eq:Hreal}).

	\section{Quadrature representation}
	Recall that the fermionic Kitaev chain is best understood in terms of Hermitian Majorana fermion operators, as opposed to the original (Dirac) fermions \citep{Alicea2012}.  The Majorana operators are defined via $\hc_j = (\hat{\gamma}^A_j + i \hat{\gamma}^B_j)/2$.  When the Hamiltonian in Eq.~(\ref{eq:HFermion}) is expressed using these operators, one finds that $A$-type Majoranas only couple to $B$-type Majoranas on neighbouring sites.  Morever, this coupling is spatially asymmetric.	This unusual pairing directly leads to the existence of unpaired Majorana end modes in an open chain.
	
	Remarkably, the bosonic model in Eq.~(\ref{eq:Hreal}) exhibits an analogous structure when we express it in terms of Hermitian quadrature operators $\hx_j$,$\hp_j$ on each site, defined via
	$\hat{a}_j = \left(\hx_j + i \hp_j \right)/\sqrt{2}$.  
	A direct substitution yields:
	\begin{equation}
	\label{eq:Hquad}
	\hH_B \equiv \frac{1}{2}
	\sum_{j}  \left(
	-(t-\Delta) \hx_{j+1} \hp_j + (t+\Delta) \hp_{j+1} \hx_j 
	\right)
	\end{equation}
	The structure here is analogous to the fermionic case: $\hx$ quadratures are only coupled to $\hp$ quadratures, and further, there is an asymmetry in the coupling between $\hx_j$ and $\hp_{j\pm1}$.   
	
	In the fermionic Kitaev chain, the system is gapped: the hopping asymmetry leads to isolated Majorana modes on the edges while the bulk can only carry supercurrent. Conversely, the asymmetric coupling in the bosonic version gives rise to unusual propagation within the bulk which has no fermionic counterpart.  Indeed, perhaps the most dramatic consequence of the quadrature pairing structure in Eq.~(\ref{eq:Hquad}) is in the dynamics.  The Heisenberg equations of motion corresponding to $\hH_B$ (with $\hbar = 1$ throughout) are:
	\begin{align}
	\label{eq:X_EOM}
	\dot{\hx}_j & = 
	\frac{t+\Delta}{2} \hx_{j-1} - \frac{t-\Delta}{2} \hx_{j+1}\\
	\label{eq:P_EOM}
	\dot{\hp}_j & = 
	\frac{t-\Delta}{2} \hp_{j-1} - \frac{t+\Delta}{2} \hp_{j+1}.  
	\end{align}
	We see that the dynamics of the $\hx$ quadratures are completely independent of the $\hp$ quadratures.  Further, each quadrature is forced in a spatially-asymmetric manner by its neighbours.  As $t$ approaches $\Delta$, we have complete asymmetry:  $\hx$ quadratures are only forced by their neighbours to the left, and $\hp$ quadratures are only forced by their neighbours to the right.  Viewing quadratures as particles, we would thus expect chiral propagation:  $x$ ``particles" would only propagate to the right, and $p$ particles only to the left.  As we will see, this basic expectation is borne out:  Eqs.~(\ref{eq:X_EOM})-(\ref{eq:P_EOM}) imply that the transport of a photonic wavepacket in our chain will be directional, with the direction determined by the {\it phase} of the wavepacket. This is reminiscent of edge states in the quantum spin Hall effect, where directionality of an edge state is determined by its spin \cite{Kane-Mele-Sep,Kane-Mele-Nov,Maciejko:2011kr}.
	
	Before proceeding, it is important to ask whether our system is dynamically stable:  can $\hH_B$ be diagonalized?  Unlike a fermionic system, the pairing terms in Eq.~(\ref{eq:Hquad}) can lead to dynamical instability, analogous to a standard parametric instability.  Focusing on a finite chain with open boundary conditions, we find that the system is stable as long as $t > \Delta$, independent of the chain length.  In the stable regime, the Hamiltonian is unitarily equivalent to a simple tight-binding chain with no parametric drive, as we now show.
	
	Defining the parameter $r$ via
	\begin{equation}
	e^{2r} = (t+\Delta)/(t-\Delta),
	\label{eq:rDefn}
	\end{equation} 
	we consider a position-dependent local squeezing transformation defined by
	\begin{align}  
	\hat{U} \hx_j \hat{U}^\dagger = e^{r (j-j_0)}\htx_j ,
	\hspace{1 cm}
	\hat{U} \hp_j \hat{U}^\dagger   
	= e^{-r (j-j_0)} \htp_j.
	\label{eq:SqzTrans}
	\end{align}
	Here $\htx_j,\htp_j$ are new canonical quadratures, and $j_0$ is an arbitrary real number.  
	One finds that
	\begin{align}
	\label{eq:RegTB}
	\hat{U} \hH_B \hat{U}^\dagger & = 
	\frac{\tilde{t}}{2}
	\sum_{j}  \left(
	-\htx_{j+1} \htp_j + \htp_{j+1} \htx_j 
	\right) 
	\nonumber        \\
	& = 
	\frac{1}{2} \sum_{j }
	\left( 
	i \tilde{t}  \hta^{\dagger}_{j+1} \hta_j + h.c. \right),
	\end{align}
	with
	\begin{equation}
	\tilde{t} = \sqrt{t^2 - \Delta^2}
	= t / \cosh(r).
	\label{eq:ttilde}
	\end{equation}
	Thus, in the stable regime $t > \Delta$, the system is unitarily equivalent to a a simple (excitation-conserving) tight-binding chain with a hopping matrix element $i\tilde{t}$. Note that the transformed Hamiltonian is completely independent of the parameter $j_0$.  This reflects the invariance of $\hat{H}_{B}$ under any spatially uniform squeezing transformation that does not mix $\hat{x}$ and $\hat{p}$ quadratures, i.e.~ $\hat{x}_j \rightarrow e^z \hat{x}_j$,  $\hat{p}_j \rightarrow e^{-z} \hat{p}_j$.
	
	We also briefly comment on the special case $t = \Delta$.  In this case, Eq.~(\ref{eq:Hquad}) implies a complete coupling asymmetry:  $\hx_j$ is {\it only} coupled to $\hp_{j+1}$ (and not to $\hp_{j-1}$).  For an open chain with $N$ sites, there are thus two localized  quadrature operators at the ends of the chain, $\hp_1$ and $\hx_N$, that drop out of the Hamiltonian.  While in the fermionic model, the corresponding decoupled modes are of great interest, they are of less interest here.  Unlike with fermions, these decoupled quadrature operators cannot be recombined to form a new, delocalized canonical bosonic mode. While these decoupled quadratures do represent quantum non-demolition (QND) observables, their existence here requires precisely tuning to the threshold of instability.

	\begin{figure*}
		\centering
		\includegraphics[width=1\textwidth]{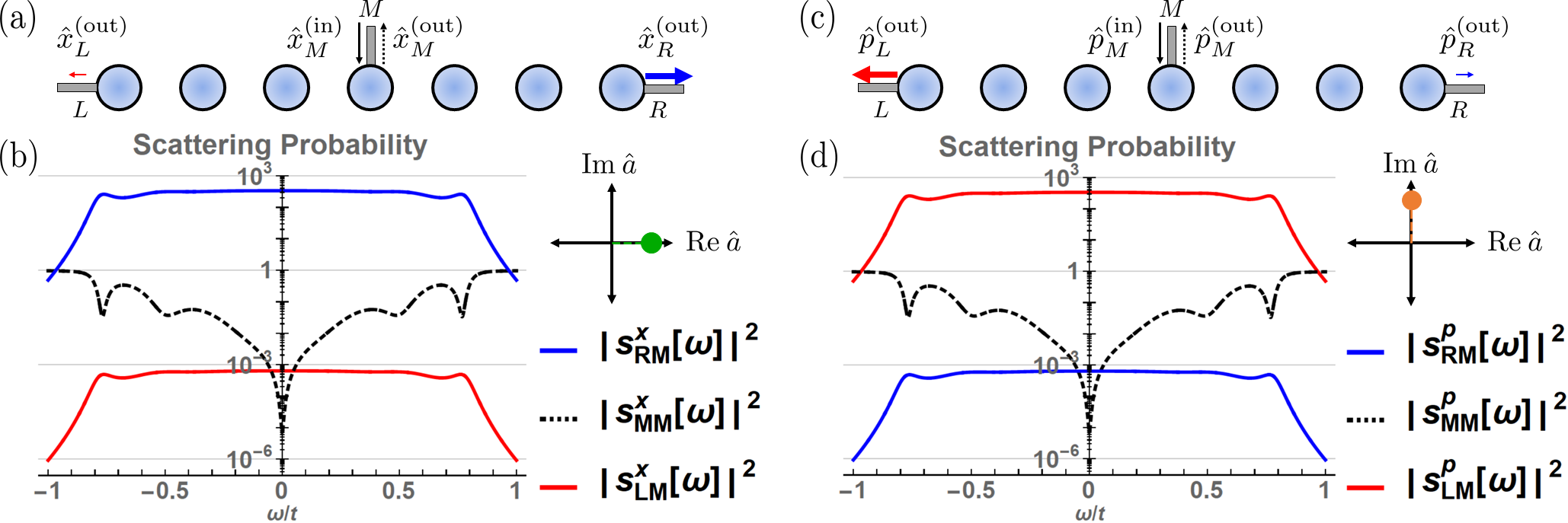} 
		\caption{Scattering properties of the bosonic Kitaev-Majorana chain: (a) Schematic of the setup. The leftmost, middle and rightmost sites are attached to waveguides (coupling rates $\kappa_L, \kappa_M$ and $\kappa_R$ respectively). A signal with a frequency $\omega$ and global phase $\theta = 0$, corresponding to an $x$ excitation, is injected in the middle of the middle waveguide and is amplified (deamplified) as it propagates to the right (left). (b) Amplitude squared of the scattering matrix elements plotted as a function of frequency of the input signal. As expected, signals propagating to the right (left) and amplified (deamplified). Note the reflection probability (black) is bounded by unity. (c) Same setup as in (a), except the phase of the signal is now $\theta = \frac{\pi}{2}$, corresponding to a $p$ excitation. (d) The signal is now amplified (deamplified) as it propagates to the left (right). 
			For (b),(d), we take $N = 13$ sites, $\Delta = t/2$, uniform on-site internal loss rate $\kappa = 10^{-2}t$, and waveguide couplings $\kappa_M = 2\kappa_L = 2 \kappa_R = 2t$. }
		\label{fig:scattering}
	\end{figure*}
	\section{Phase-sensitive chiral propagation}
	Eqs.~(\ref{eq:X_EOM}),(\ref{eq:P_EOM}) imply that photonic excitations in our lattice propagate in a chiral fashion, with a direction that depends on the {\it phase} of the excitation.  To fully characterize this behaviour, we calculate the retarded single-particle Green's functions for our system in position space, focusing on an $N$-site open chain.  To make the chirality explicit, we calculate quadrature-quadrature Green's functions.  This is easily done using the position-dependent squeezing transformation in Eq.~(\ref{eq:SqzTrans}).   
	One finds that the only non-zero Green's functions are:
	\begin{align}
	G^R_{x}[j,j';\omega] & 
	\equiv -i \int_0^\infty dt \, e^{i \omega t} \langle [ \hx_j(t), \hp_{j'}(0)] \rangle  \nonumber \\
	& = i \G0[j,j';\omega] e^{r(j-j')} 
	\label{eq:GRxx}  \\
	G^R_{p}[j,j';\omega]  
	& = -i \G0[j,j';\omega] e^{-r(j-j')} \label{eq:GRpp}
	\end{align}
	Here, $\G0[j,j';\omega]$ is the retarded photon Green's function of an $N$-site tight-binding chain with tunnel matrix element $i \tilde{t}$ (see Appendix \ref{app:TBChain}).  
	
	The Green's function $G^R_{x}[j,j'; \omega]$ describes propagation of the $x$ quadrature in the lattice; more explicitly, it describes how $\hat{x}_j$ responds to a perturbation which directly forces $\hat{x}_{j'}$.  $G^R_{p}[j,j'; \omega]$ is interpreted analogously.    The Green's functions above directly manifest the expected chirality:  $x$ quadrature signals are amplified (de-amplified) as they propagate from left to right (right to left); the $p$ quadratures exhibit the {\it opposite} behaviour. 
	Note that the local Green's functions ($j=j'$) have no explicit $r$ dependence or phase-sensitivity:  they are identical to those of a particle-conserving tight-binding model with hopping $i\tilde{t}$.
	
	Remarkably, the above structure still holds if we break translational invariance by introducing position-dependent loss on each site.  We treat this loss as Markovian, and model it using standard input-output theory \cite{Gardiner00,ClerkRMP}. The loss gives rise to a damping rate $\kappa_j$ on each lattice site; 
	it could be due to a deliberate coupling to waveguides, or to internal loss.  The Heisenberg-Langevin equations for the $\hx$ and $\hp$ quadratures on each site now read
	\begin{align}
	\label{eq:LangevinX}
	\dot{\hx}_j = 
	\frac{\tilde{t}}{2}
	\left(
	e^{r} \hx_{j-1} - e^{-r}\hx_{j+1} \right)
	-{1\over 2} \kappa_j \hx_j-\sqrt{\kappa_j}\hx^{\text{(in)}}_j, \\
	\dot{\hp}_j = 
	\frac{\tilde{t}}{2}
	\left( 
	e^{-r} \hp_{j-1}
	-e^{r}\hp_{j+1} \right) 
	-{1\over 2}\kappa_j \hp_j-\sqrt{\kappa_j}\hp^{\text{(in)}}_j.
	\label{eq:LangevinP}
	\end{align}
	$\hx^{\text{(in)}}_j(t)$ and $\hp^{\text{(in)}}_j(t)$ describe the quadratures of the input field associated with the loss port $\kappa_j$; in the simplest case, they just describe vacuum fluctuations.  
	
	Despite the additional terms due to loss, we can still map our system to a simple tight-binding model with no pairing terms.  One simply applies the local site dependent squeezing transformation in Eq.~(\ref{eq:SqzTrans}) to the Langevin equations.  After this transformation, the Heisenberg-Langevin equations describe a simple tight-binding model with renormalized hopping $\tilde{t}$ (c.f.~Eq.~(\ref{eq:ttilde})) and decay rate $\kappa_j$ on each site.  While the input noise operators now describe squeezed noise, the linearity of the equations means that this does not have any effect on the Green's functions.  As a result, even with loss, the Green's functions of our system still have the form of Eqs.~(\ref{eq:GRxx}-\ref{eq:GRpp}), where now $\G0[j,j';\omega]$ is the photonic Green function of a tight-binding chain with hopping $\tilde{t}$ and on-site loss rates $\kappa_j$
	(see Appendix \ref{app:TBChain}).
	
	Finally, we stress that the phase-dependent chirality manifested by the Green functions also reflects itself in simple wavepacket dynamics.  
	To be concrete, suppose we initially prepare our lattice in a coherent state wavepacket with zero average momentum .  Such a state is characterized by $\langle \hat{a}_j(0) \rangle = e^{i \theta} f_j$, where $\theta$ is the \textit{global phase} of the excitation and $f_j>0$ describes the envelope of the wavepacket.  We can now directly use Eqs.~(\ref{eq:X_EOM}),(\ref{eq:P_EOM}), to determine the evolution of $\langle \hat{a}_j(t) \rangle$ and hence the motion of the wavepacket.  We see that the wavepacket will split into two:  the $\cos \theta$ component of the wavepacket corresponds to an $x$ quadrature excitation that propagates and is amplified as it moves to the right, while the $\sin \theta$ component is a $p$ quadrature excitation that propagates to the left.
  
	\section{Scattering Properties}
	We now consider the case where input-output waveguides are attached to our lattice, and ask how fields incident on the lattice from these waveguides are scattered.  The relevant scattering matrix follows immediately from the input-output boundary condition $\hat{a}^{(\rm out)}_j = 
	\hat{a}^{(\rm in)}_j + \sqrt{\kappa_j} \hat{a}_j$ \cite{Gardiner00,ClerkRMP}
	and the Heisenberg-Langevin equations in Eqs.(\ref{eq:LangevinX}),(\ref{eq:LangevinP}).  As $x$ and $p$ quadratures are dynamically decoupled in our system, scattering does not mix these quadratures.  The scattering matrix thus takes a simple form in the quadrature basis, and is defined by
	\begin{align}
	\hx_{j}^{\text{(out)}}[\omega] & = \sum_{j'} s_{jj'}^{x}[\omega] \hx_{j'}^{\text{(in)}}[\omega], 
	\nonumber \\
	\hp_{j}^{\text{(out)}}[\omega] & = \sum_{j'}s_{jj'}^{p}[\omega] \,\hp_{j'}^{\text{(in)}}[\omega]
	\end{align}
	The scattering matrix is directly determined by the system Green's functions, and thus inherits their structure:
	\begin{align}
	\label{eq:Sxx_Matrix}
	s_{jj'}^{x}[\omega]  = e^{r(j-j')} \ts_{jj'}[\omega],
	\\
	\label{eq:Spp_Matrix}
	s_{jj'}^{p}[\omega] = e^{-r(j-j')} \ts_{jj'}[\omega].
	\end{align}
	Here $\ts_{jj'}[\omega]$ is the scattering matrix of a $N$-site tight-binding chain with tunnel matrix element $i\tilde{t}$ and decay rate $\kappa_j$ on each site (see Appendix \ref{app:TBChain}); the scattering in such a system is insensitive to phase, and hence the same for any quadrature. We see that in our full system, as expected, the $\hx$ quadrature is amplified (deamplified) in transmission from left to right (right to left) whereas the $\hp$ quadrature exhibits the opposite behavior (see Fig.~\ref{fig:scattering}). 
	
	Our system thus represents a unique kind of phase-sensitive amplifier.  Such devices amplify only one quadrature of an input signal (deamplifying the other conjugate quadrature), and are capable of quantum amplification with zero added noise \cite{ClerkRMP}.
	They also serve as sources of non-classical quadrature squeezed light. Standard phase sensitive amplifiers either only amplify in one direction (due to phase matching), or are reciprocal, and amplify the same quadrature irrespective of transmission direction. In contrast, we obtain amplification in both directions, but in different orthogonal quadratures.
	Note that the end-to-end transmission gains $|s_{N1}^{x}[\omega]|^2 = |s_{1N}^{p}[\omega]|^2$ scale like $e^{2rN}$, while the amplification bandwidth scales like $\tilde{t} = t / \cosh(r)  $.  Our system is thus not limited by a standard gain-bandwidth product: by using a long chain and a small $r$, large gain is possible without sacrificing bandwidth.
	
	Viewed as an amplifier, our system has another remarkable property:  while there is strong gain in transmission, there is {\it never} any gain in reflection.  This follows immediately from Eqs.(\ref{eq:Sxx_Matrix}),(\ref{eq:Spp_Matrix}), which tells us that 
	$s_{jj}^{x}[\omega] = s_{jj}^{p}[\omega] = \ts_{jj}[\omega]$: the reflection amplitude coincides with that of a simple tight-binding model with hopping $i\tilde{t}$, and hence can never be larger than unity.  
	The lack of reflection gain is of practical utility in many settings, where one wishes to protect a fragile signal source coupled to an input port.  
	
	For an intuitive picture of the lack of reflection gain, consider trajectories and the equations of motion in Eqs.~(\ref{eq:X_EOM})-(\ref{eq:P_EOM}). Any reflection process requires an equal amount of left-to-right propagation and right-to-left propagation.  In our system, the directional nature of the amplification means that the net amplification for such a process will always be zero:  amplification in one direction is perfectly compensated by de-amplification in the opposite direction.   	
	
	\section{Extreme sensitivity to boundary conditions}
	
	\begin{figure}[!t]
		\centering
		\includegraphics[width=0.45\textwidth]{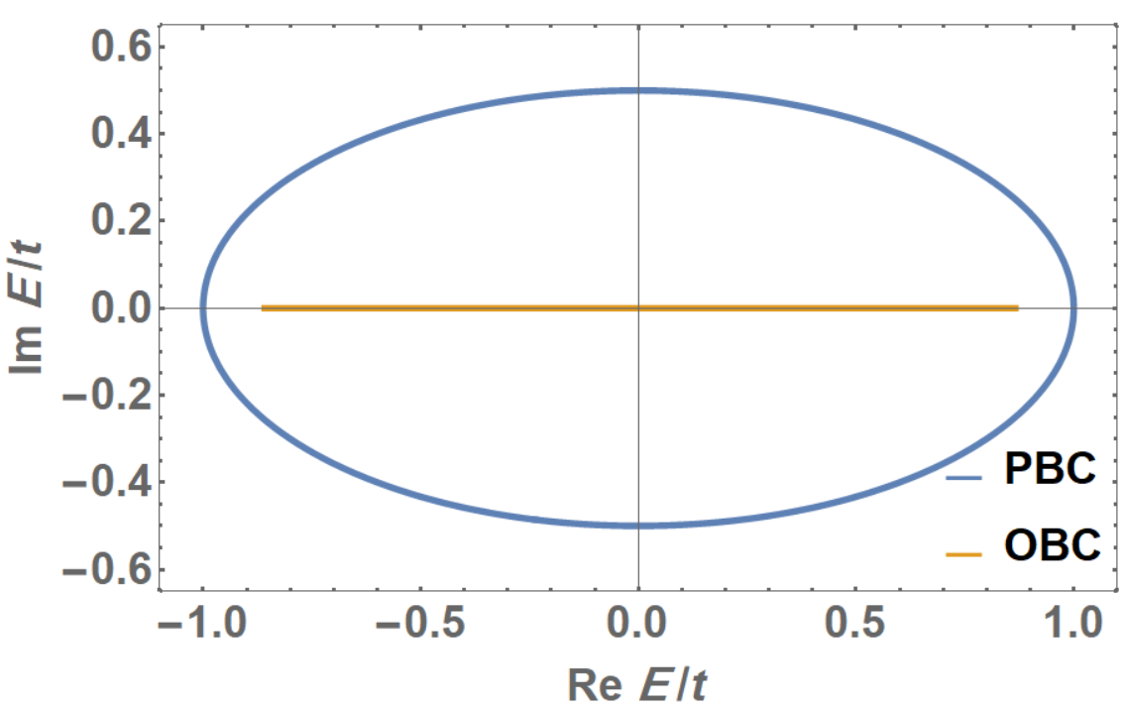}
		\caption{Spectrum of the system with periodic boundary condition (PBC) $E^{PBC}_{k,\pm} = t \sin(k)\pm i \Delta \cos(k)$ versus open boundary conditions (OBC) $E_k^{OBC} = \sqrt{t^2-\Delta^2}\cos(k)$ with $\Delta = t/2$. The spectrum with PBC is complex for any non-zero $\Delta$, indicating parametric instability. In contrast, the system with OBC is stable as long as $t> \Delta$, regardless of system size. } 
		\label{fig:spectrum}
	\end{figure}

	A key feature of the fermionic Kitaev-Majorana chain is a marked sensitivity to boundary conditions:  for periodic boundary conditions, the system has an energy gap centered around zero energy, whereas with open boundary conditions, there exist localized, zero energy Majorana edge modes.  We find that our bosonic analogue of the Kitaev chain also exhibits an extreme sensitivity to boundary conditions:  with periodic boundary conditions, the system is {\it always} characterized by unstable, spatially extended eigenmodes, whereas with open boundary conditions it can be completely stable (all mode energies real), and further, have completely localized wavefunctions.  We explain this further in what follows.  
	
	Consider first the system with periodic boundary conditions and no dissipation.  The Heisenberg equations of motion in momentum space take the form
	\begin{equation}
	\left(i \partial_t - \check{M}_k \right)
	\begin{pmatrix}
	\ha_k \\ \ha^{\dagger}_{-k}
	\end{pmatrix} 
	\hspace{0.2 cm}
	\check{M}_k =
	t \sin(k) \check{1} + i \Delta \cos(k) \check{\sigma}_x
	\label{eq:MkDefn}
	\end{equation}
	As usual, the mode energies are the eigenvalues of the dynamical matrix $\check{M}_k$, and are given by $E_{k,\pm} = t \sin(k)\pm i \Delta \cos(k)$.  The fact that the eigenvalues are complex for all $k$ (except $\pm \pi/2$) indicates that for {\it any} nonzero $\Delta$, the system is past the threshold for parametric instability, and will exhibit exponential growth in the time domain.  This is not surprising as the parametric drive is resonant:  the drive is adding pairs of photons with zero total momentum, and the energy detuning of such a pair is always zero.  Note that the wavefunctions of these modes are simple planewaves, consistent with the translational invariance of the model.   
	
	While intuitively reasonable, the behaviour of the periodic boundary condition system is in stark contrast to the open-boundary condition system.  As already shown in Eq.~(\ref{eq:SqzTrans}), in this case we can make a unitary squeezing transformation for any $\Delta < t$ to map our system to a simple tight-binding model with hopping matrix element $i\tilde{t} = i\sqrt{t^2 - \Delta^2}$.  As such, there is no instability:  all mode eigenvalues are real, given by $E_{n} =  \tilde{t} \cos k_n$, with $k_n = n \pi / (N+1)$.  We thus have a dramatic difference in the spectrum of the model depending on the choice of boundary conditions (see Fig.~$\ref{fig:spectrum}$).  Note that this conclusion is independent of system size. 
	
	The difference between a ring and chain geometry goes beyond just the spectrum: the wavefunctions are also completely different in the two cases.  Naively, one would expect that for the open chain, the eigenstates are simple standing waves, formed by taking linear combinations of the plane wave eigenstates of the ring.  This is not the case.  For an $N$-site open chain, the diagonalized Hamiltonian can be written as $\hat{H}_{B}^{OBC} = \sum_n E_{n} \hat{\beta}^\dagger_n \hat{\beta}_n$.  The quasiparticle $\hat{\beta}_n$ are given by a Bogliubov transformation of our original real-space photon operators,  
	\begin{align}
	\hat{\beta}_n = \sum_{j=1}^N \left[ u_n(j)\hat{a}_j-v_n(j)\hat{a}^\dagger_j \right],
	\end{align}
	with the functions $u_n(j)$, $v_n(j)$ representing the ``particle" and ``anti-particle" parts of the wavefunction.  Using the squeezing transformation in Eq.~(\ref{eq:SqzTrans}) which diagonalizes our Hamiltonian, one finds easily
	\begin{align}
	\label{eq:EigenParticle}
	&u_n(j) = \sqrt{\frac{2}{N+1}}i^{-j} 
	\sin (k_n j) \cosh(r(j-j_0)) \\
	\label{eq:EigenHole}
	&v_n(j) = \sqrt{\frac{2}{N+1}}i^{-j} 
	\sin (k_n j) \sinh(r(j-j_0)). 
	\end{align}
	where the squeeze parameter $r$ is defined in Eq.~(\ref{eq:rDefn}). Note that the chiral symmetry of our system implies that there are many possible choices of eigenmode basis; this corresponds to the freedom to pick the parameter $j_0$ defining our squeezing transformation.
	
	We see that the particle and anti-particle parts of the mode wavefunction are each localized:  they have an exponential dependence on position, and their weight is concentrated at the ends of the chain. On the other hand, for a given eigenmode, the total contribution of each site to the symplectic norm $|u_n(j)|^2-|v_n(j)|^2$ does not exhibit any sort of localization.  This quantity is in fact completely independent of the two-photon driving amplitude $\Delta$. This is consistent with the local Green’s function being phase-insensitive and independent of $r$.  The upshot is that it is difficult to detect the localization of mode wavefunctions in our system using purely local probes; one must instead consider a non-local probe such as transmission.
	
\begin{figure}[t]
		\centering
		\includegraphics[width=0.47\textwidth]{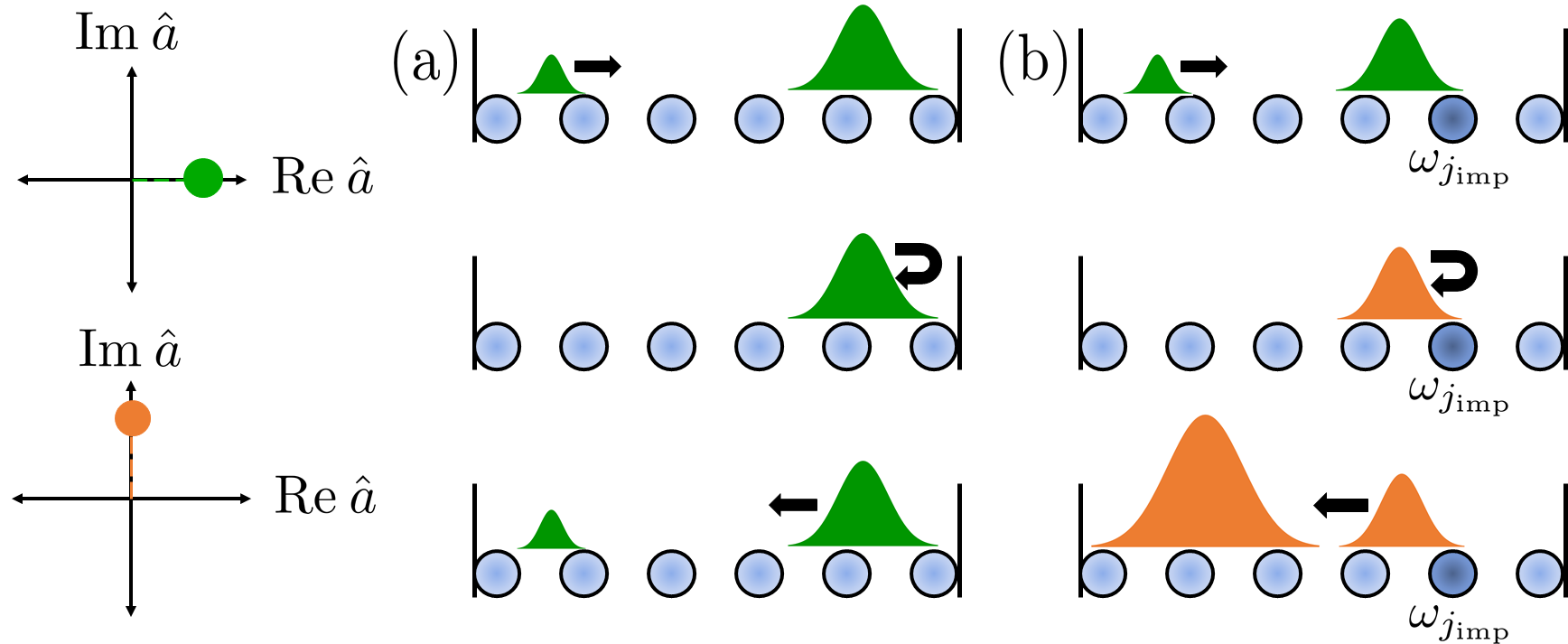}
		\caption{(a) An $x$ excitation is amplified as it propagates to the right. After being reflected off the boundary, it remains an $x$ excitation as propagates to the left while being deamplified. The scattering mechanism is the same if we  we consider reflecting off of on-site loss.  (b) An $x$ excitation can turn into a $p$ excitation after scattering off an impurity (dark blue site). In combination with the chiral nature of amplification, multiple reflections between impurities can lead to instability.} 
		\label{fig:ImpurityScattering}
	\end{figure}

	For an intuitive picture of our model's  striking sensitivity to boundary conditions,  we return to our picture of chiral wavepacket dynamics.  Recall that e.g.~an $x$ quadrature excitation propagates to the right with amplification, and to the left with de-amplification.  Thus, in a ring geometry, an initial $x$ excitation can propagate and be amplified indefinitely as it traverses the ring several times in a clockwise direction.  This infinite amplification directly leads to dynamical instability.  In contrast, in a finite chain, the disturbance will eventually hit the right boundary of the system and be reflected.  As it now propagates to the left, it will be de-amplified.  There is thus no possibility for indefinite amplification, and the system remains stable (see Fig.~\ref{fig:ImpurityScattering}).  
	\section{Disorder effects}
	We now explore the sensitivity of our lattice to various types of disorder, focusing on the case of an open chain with $N$ sites. The simplest kind of disorder in our model is random on-site losses $\kappa_j$.  We have already seen that such disorder is innocuous: while random loss breaks translational invariance causes reflections within the chain, it never causes any instability:  the system remains unitarily equivalent to a simple tight binding chain with disordered loss, c.f.~Eqs.~(\ref{eq:LangevinX})-(\ref{eq:LangevinP}).  
	
	The situation is markedly different if we have non-zero on-site energies, described by:
	\begin{equation}
	\hat{H}_{\rm dis} = \sum_{j=1}^N \omega_j \hat{a}^\dagger_j \hat{a}_j
	\label{eq:Hdis}
	\end{equation}
	Such potential disorder can induce instability in our system even if $t < \Delta$.  Formally, this can be understood from the local squeezing transformation in Eq.~(\ref{eq:SqzTrans}) that maps our system onto a simple tight-binding chain.  While on-site loss terms are invariant under this mapping, on-site potential terms transform to particle non-conserving parametric drive terms, i.e.:
	\begin{eqnarray}
	\hat{U}\hat{H}_{\rm dis} \hat{U}^\dagger & = 	
	\sum_{j=1}^N \omega_j 
	\Big(
	\cosh(2 r (j-j_0)) 
	\hat{\tilde{a}}^\dagger_j \hat{\tilde{a}}_j 
	\nonumber \\
	& + \frac{1}{2} \sinh(2 r (j-j_0)) \left[
	\hat{\tilde{a}}^\dagger_j \hat{\tilde{a}}^\dagger_j
	+\hat{\tilde{a}}_j \hat{\tilde{a}}_j
	\right] \Big)
	\end{eqnarray}
	where we've thrown away terms proportional to the identity. It is thus generically impossible to transform to a frame where our system conserves particle number, and thus one can generically get dynamical instability.  Note that in the special case where there is just a single impurity at site $j=j_{\rm imp}$, the system is still always stable: formally, we could pick the origin $j_0$ of our squeezing transformation to coincide with $j_{\rm imp}$, and thus map our system onto a particle conserving model. This is no longer the case if we were to add even one additional impurity at any other lattice site. 
		
	For a more heuristic understanding of how impurities cause instability, we return to the equations of motion for local quadratures.  With disorder, they now read:
	\begin{align}
	\dot{\hx}_j =
	\frac{\tilde{t}}{2}
	\left(
	e^{r} \hx_{j-1} - e^{-r}\hx_{j+1} \right)
	-\frac{1}{2}\kappa_j\hx_j+\omega_j\hp_j,\\
	\dot{\hp}_j =
	\frac{\tilde{t}}{2}
	\left(
	e^{-r}\hp_{j-1}
	-e^{r} \hp_{j+1} \right)
	-\frac{1}{2}\kappa_j\hp_j-\omega_j\hx_j.
	\end{align}
	We see that while loss never dynamically couples $x$ and $p$ quadratures, the same is not true of frequency disorder.  Scattering off impurities can now change the $\it{phase}$ of an excitation, (i.e.~convert $x$ to $p$ and vice-versa).  The chiral nature of amplification in our lattice implies that multiple scatterings of this type can lead to indefinite amplification and thus instability (see Fig.~\ref{fig:ImpurityScattering}). This is in contrast to scattering off of local on-site loss or off the boundaries of a finite-sized system, processes which manifestly preserves the phase of excitation and hence do not lead to any instability. 
	
\begin{figure}[t]
		\centering    \includegraphics[width=0.45\textwidth]{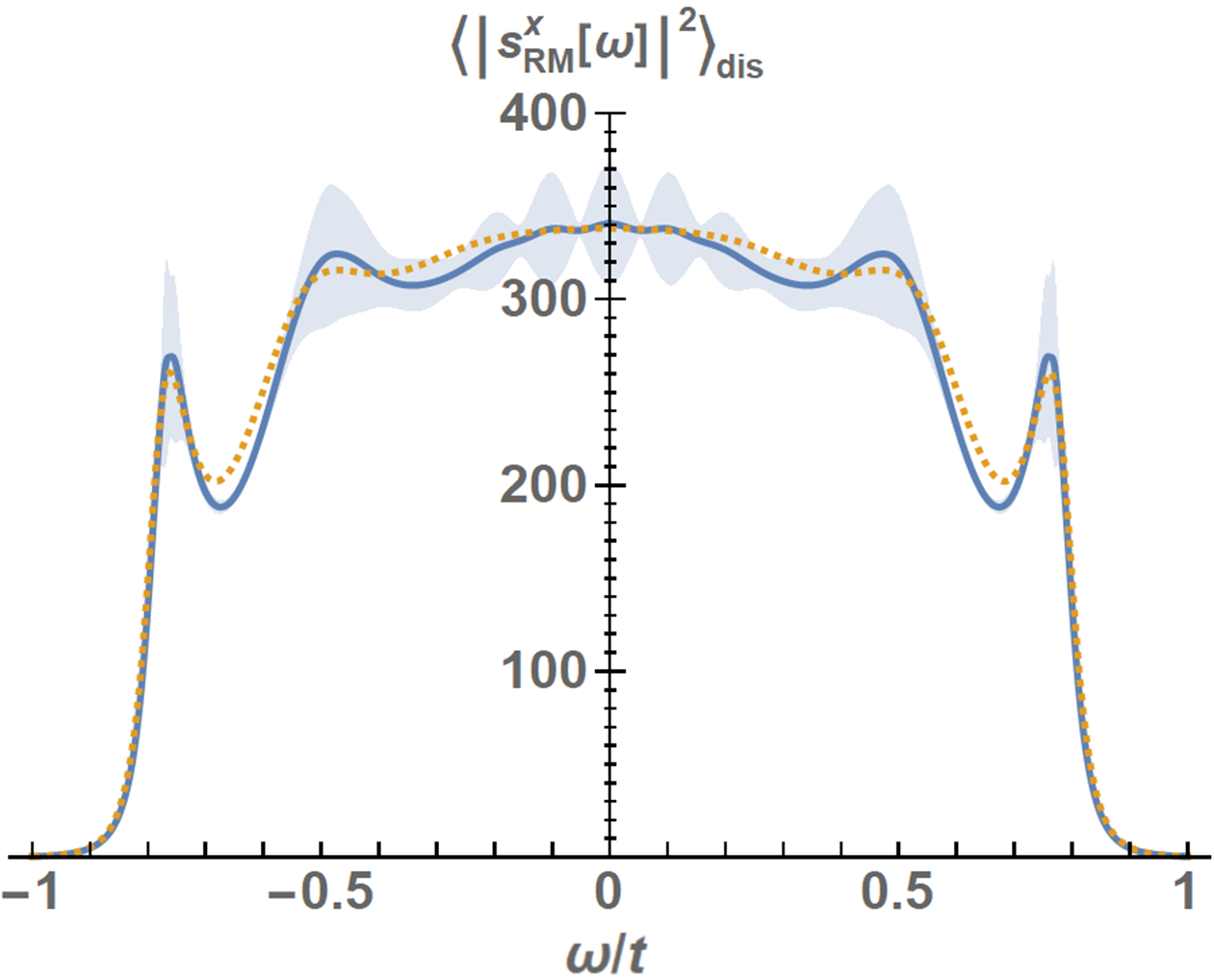}
		\caption{Disorder averaged transmission coefficient $\langle |s_{\rm{RM}}^{x}[\omega]|^2 \rangle_{\rm{dis}}$  for the same setup as is Fig.~\ref{fig:scattering}, but with on-site disorder (disorder strength $W = 10^{-3} t$). The shaded region corresponds to the variance and the dashed orange line corresponds to the clean system. Although stability is no longer guaranteed after introducing the disorder potential, with the chosen parameters, instability only occurs in less than 0.01\% of realizations. For smaller values of $r$ and/or $N$, one can tolerate even larger amounts of disorder. Parameters: $N = 13$ sites, $\Delta = t/2$, uniform on-site internal loss $\kappa = 10^{-2}t$, waveguide coupling rates $\kappa_M = 2\kappa_L = 2 \kappa_R = 2t$. } 
		\label{fig:Disorder}
	\end{figure}

	To quantitatively assess the impact of frequency disorder, we numerically perform a disorder average.  We take the $\omega_j$ in Eq.~(\ref{eq:Hdis}) to be independent random variables drawn from a uniform distribution on the interval $[-W,W]$, with $W$ representing the disorder strength.  While multiple scatterings can lead to instability if there is significant amplification, disorder effects should be weak if $W \ll  \tilde{t}$ and $r,N$ are not too large.
	This is borne out by our simulations.  Fig~\ref{fig:Disorder} shows results for disorder-averaged scattering probabilities ($10^4$ realizations) for a chain analogous to that in Fig.~\ref{fig:scattering} ($N=13$, $r\approx 0.55$), but with disorder strength $W = 10^{-3} t$.  For these parameters, less than $0.01\%$ of disorder realizations yield instability.  In the remaining instances, the scattering closely resembles the behaviour of the clean system. 	
	
	\section{Generation of multi-partite entanglement}
	We now consider our system's ability to emit entangled photons. Consider the case where $M \leq N$ sites of the lattice are coupled to input-output waveguides.  Even if simple vacuum noise is incident in each of these channels, the outgoing light will have non-zero photon number, and will exhibit entanglement.  
	
\begin{figure}[t]
		\centering
		\includegraphics[width=0.45\textwidth]{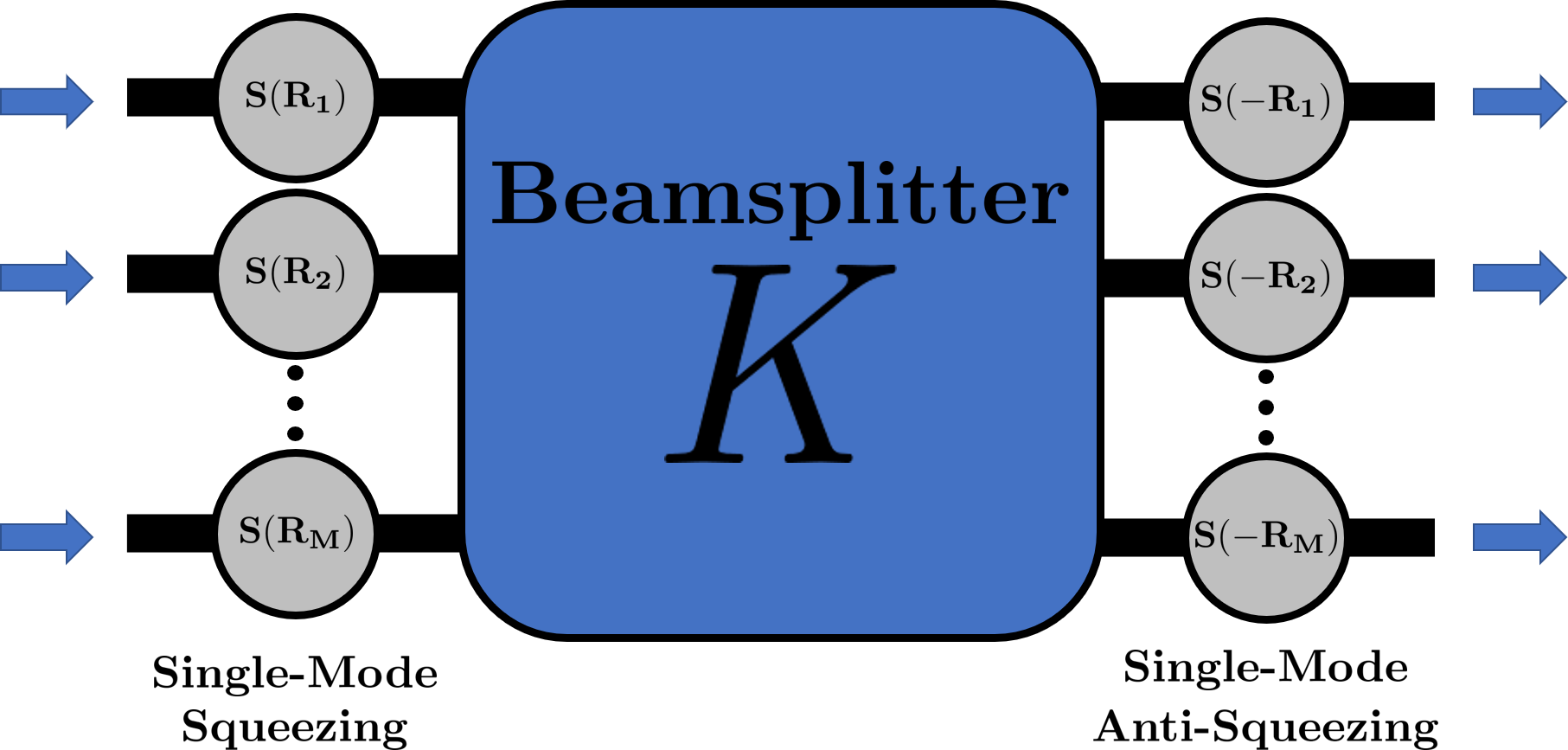}
		\caption{Schematic showing an equivalent depiction of scattering off our lattice when there are $M$ coupling waveguides.  Input states in each waveguide are first locally squeezed (squeeze parameters $R_j$).  They then pass through a beam-splitter network, and are then finally locally anti-squeezed at the outputs.  Crucially, the squeeze parameters $R_j$ and beam-splitter unitary matrix $K$ have a direct and simple relation to the system Hamiltonian.  In particular, $K$ is the unitary of a simple tight-binding model (see Eqs.~(\ref{eq:Sxx_Matrix_Disorder}-\ref{eq:Spp_Matrix_Disorder})). } 
		\label{fig:ScatteringEquiv}
	\end{figure}

	While entanglement generation is generic to any sort of bosonic quantum amplifier, our system has unique features. 
	In particular, the state of the output light has a remarkably transparent form:  it is identical to
	sending a product of single mode squeezed states into a beam-splitter network, with the squeezing parameters and beam-splitter unitary being directly determined (in a simple way) by the Hamiltonian.  This is depicted in  Fig.~\ref{fig:ScatteringEquiv}, and follows immediately from Eqs~(\ref{eq:Sxx_Matrix})-(\ref{eq:Spp_Matrix}).  
	First, one prepares $M$ single mode squeezed states (one each channel), with squeeze parameter $R_j = r j$ (c.f.~Eq.(\ref{eq:rDefn})) in the waveguide coupled to site $j$.  This populates each channel with photons, but does not create correlations.  Next, the resulting state is sent into an effective beam-splitter network described by a $M \times M$ unitary matrix $K$; this is just the scattering matrix of a (particle-conserving) tight-binding model with hopping $i \tilde{t}$ (c.f.~Eq.~(\ref{eq:ttilde})) and waveguide couplings $\kappa_j$.  Finally, we apply another set of local, single-mode squeezing transformations on each channel.  As this is a product of purely local unitaries, this last step does not modify entanglement properties.
	
	Note that the resource complexity of such states can be quantified by the number of independent single mode squeezed states required for their production \cite{Braunstein2005}; in our case, this is $M$ and can be  extremely large.  The standard method for producing them requires the experimentally challenging task of first preparing $M$ squeezed states, then transporting and injecting them with high efficiency into a complex beam-splitter network.  Our system allows one to circumvent these difficulties by having all the required squeezing generated locally.
	
	States of the form depicted in Fig.~\ref{fig:ScatteringEquiv} have received considerable recent attention.  They are of interest as a means to demonstrate ``quantum supremacy'' \cite{Aaronson2013}.  Computing their photon statistics (i.e.~Gaussian boson sampling \cite{Hamilton2017GBS}) requires calculating the halfnian of a $M \times M$ matrix, something that is known to be computationally hard classically (being in the $\sharp P$ complexity class). Output states of this form are also a resource for simulating molecular vibronic spectra \cite{Huh2015GBS}, and for solving certain classically-hard graph-theoretic problems \cite{2017arXiv171206729B}.  
	
	For applications, one ideally wants the ability realize a variety of beam-splitter operations $K$ appearing in Fig.~\ref{fig:ScatteringEquiv}.  The requisite tunability can be achieved by allowing the magnitude of the hopping and parametric driving on each bond to vary, i.e.~$t \rightarrow t_j$, $\Delta \rightarrow \Delta_j$.  As shown in Appendix \ref{app:Generalized Model}, as long as $\Delta_j < t_j$ on each bond, our system remains dynamically stable, and the scattering continues to correspond to the schematic in Fig.~\ref{fig:ScatteringEquiv}.  Now however the beam-splitter unitary $K$ corresponds to the scattering matrix of a non-uniform tight-binding chain with hoppings $\tilde{t}_j = \sqrt{t_j^2 - \Delta_j^2}$.  Thus, in realizations of our model where one can control $t_j, \Delta_j$, one has the ability to realize a wide class of non-trivial multi-mode entangled output states.


	\section{Connections to non-Hermitian models}
	It is natural to ask whether the striking features of our bosonic model can be given a topological underpinning.  To address this, we first make the simple observation that while our Hamiltonian is clearly Hermitian, the mode eigenvalues follow from diagonalizing the system's dynamical matrix (e.g.~$\check{M}_k$ in momentum space, c.f.~Eq.~(\ref{eq:MkDefn})).  This matrix is explicitly {\it non-Hermitian}.  As such, there is an intimate connection between our model (and parametrically-driven bosonic systems in general) and the study of non-Hermitian quantum models.  In particular, recently developed topological invariants for non-Hermitian systems can be directly applied to our system.  
	
	Note that the dynamical matrix $\check{M}_k$ (c.f.~Eq.~(\ref{eq:MkDefn})) of our model is unitarily equivalent to the non-Hermitian Hamiltonian:
	\begin{align}	
	\check{H}_{\rm HN} & = 
	t \sin(k) \check{1} + i \Delta \cos(k) \check{\sigma}_z 
	\nonumber \\
	& = 
	\frac{-i}{2} \begin{pmatrix}
	t_- e^{ik} - t_+ e^{-ik} 
	& 0 \\
	0 & 
	t_+ e^{ik} - t_- e^{-ik}
	\end{pmatrix}
	\label{eq:NonHermitian}
	\end{align}
	where $t_{\pm} \equiv t \pm \Delta$.
	This non-Hermitian Hamiltonian describes two decoupled 1D chains with asymmetric nearest neighbour hopping, i.e.~two copies of the well-known Hatano-Nelson model \cite{Hatano1997}. One chain has stronger left-to-right hopping, while the other has stronger right-to-left hopping.  These two effective chains correspond directly to the chirality of quadrature propagation as described by Eqs.(\ref{eq:X_EOM})-(\ref{eq:P_EOM}).
	
	Non-Hermitian asymmetric hopping models have been the subject of several recent studies exploring topology.  In particular, Ref.~\cite{Ueda2018a} argues that the winding of the complex spectrum of a single band, 1D, non-Hermitian model can be used to define a topological invariant, and that this invariant gives rise to a kind of bulk-boundary correspondence between an infinite system and a semi-infinite system.  This invariant could be applied to each of the decoupled 1D chains described by Eq.~(\ref{eq:NonHermitian}):  the top chain would have a winding $+1$, the bottom chain a winding $-1$.  In another study, Ref.~\cite{Xiong2017} demonstrated that non-Hermitian asymmetric hopping models exhibit striking differences in both their spectrum and wavefunctions when comparing a ring versus open chain configuration; this is also reminiscent of our model.  
	
	Non-Hermitian asymmetric hopping models are often introduced without any clear sense of how they could be physically realized; in addition, any such realization would involve dissipation, and the corresponding fluctuations could disrupt interesting behaviour.  Our work shows that parametrically driven bosonic systems give a physically-realizable platform for a class of effective non-Hermitian models, and moreover, can do this without any necessity for dissipation and noise.  
	
	\section{Physical realization}
	The parametrically-driven coupled cavity model studied in this work could be realized in a variety of different photonic and phononic systems.  $\hH_B$ in Eq.~(\ref{eq:Hreal}) describes a 1D array of tunnel-coupled cavities subject to parametric driving on each bond, where we work in a rotating frame at the parametric drive frequency (which is the same for each bond).  To construct simple physical implementations of our system, it is easiest to work in a gauge where the Hamiltonian is real, but where the pairing amplitude $\Delta$ is spatially dependent.  Making the gauge transformation 
	$\hat{a}_j \rightarrow \hat{a}_j e^{i \pi j/2}$ yields:
	\begin{equation}
	\label{eq:Hreal2}
	\hH_B = \frac{1}{2} \sum_{j }
	\left(  t  \ha^{\dagger}_{j+1} \ha_j + 
	(-1)^j \Delta  \ha^{\dagger}_{j+1} \ha_{j}^\dagger + h.c. \right).
	\end{equation}
	In this gauge, the pairing amplitude phase is modulated from site to site, corresponding to parametric driving where we inject pairs with a centre-of-mass momentum $k_{\rm tot} = \pi$.  Our system can thus be realized by using a cavity array with passive nearest-neighbour tunneling, and nearest-neighbour parametric driving with staggered phases.
	
\begin{figure}
		\centering
		\includegraphics[width=0.32\textwidth]{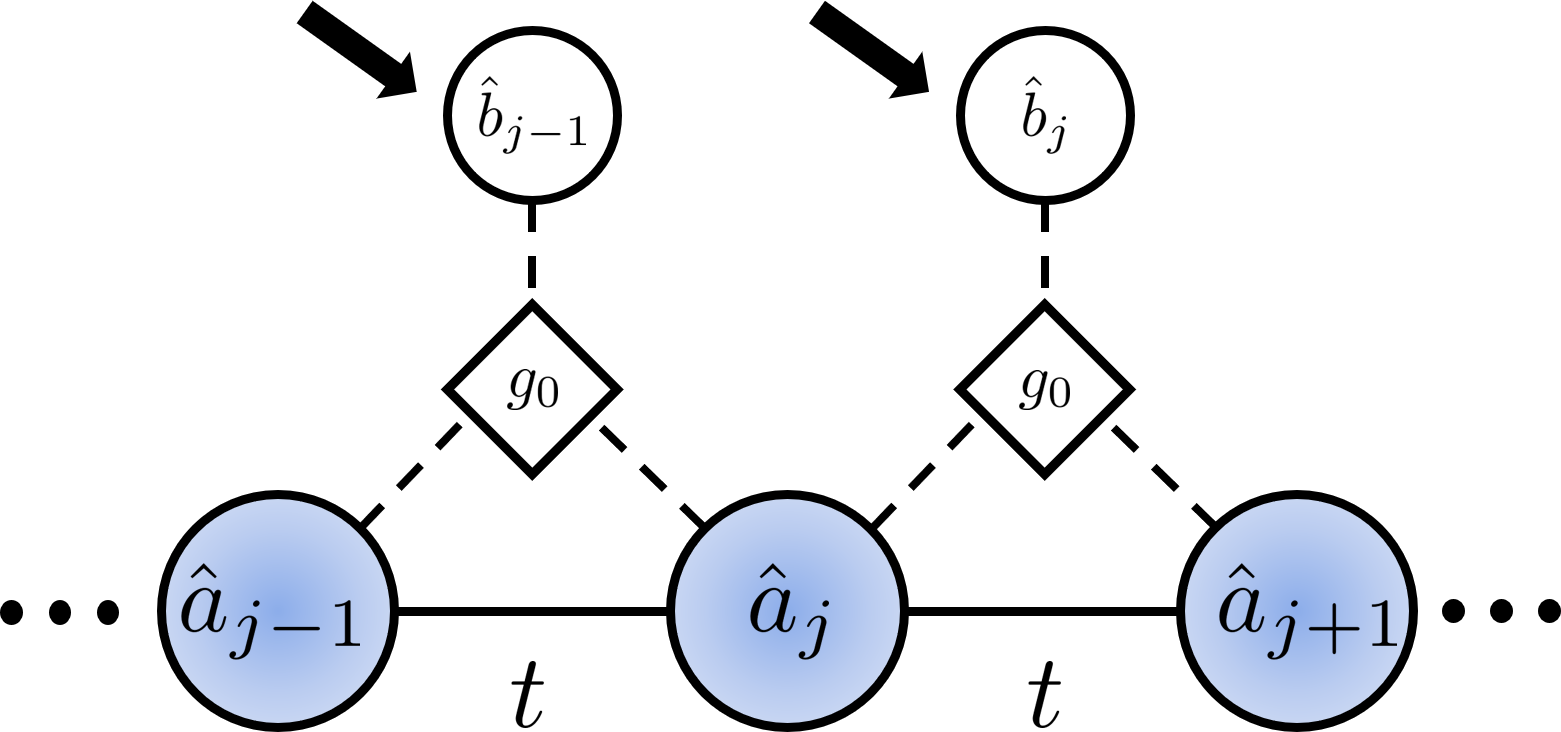}
		\caption{A possible realization of the bosonic Kiatev-Majorana chain. On each bond, adjacent cavity modes $\ha_{j}$ and $\ha_{j+1}$ are tunnel-coupled with hopping amplitude $t$. In addition, they are both coupled to a coherently driven auxiliary mode $\hat{b}_j$ via a three-wave mixing interaction (Eq.~(\ref{eq:ThreeWaveHInt})).  At the mean-field level, this interaction yields the parametric two-photon drive in our model.   Staggering the drive phases results in a Hamiltonian that is gauge-equivalent to a purely imaginary hopping phase, a key ingredient in our setup.}
		\label{fig:PhysicalRealization}
	\end{figure}

	A generic implementation is depicted schematically in Fig.~\ref{fig:PhysicalRealization}, where on each link of our main lattice, we have a nonlinear three-wave mixing interaction with an auxiliary mode
	$\hat{b}_j$: 
	\begin{equation}
	\hat{H}_{\rm int} = g_0 \sum_j 
	\left(\ha_{j+1} + \ha_{j+1}^\dagger \right)
	\left(\ha_j + \ha_j^\dagger \right)
	\left(\hat{b}_j + \hat{b}_j^\dagger \right).
	\label{eq:ThreeWaveHInt}
	\end{equation}
	By coherently driving the $\hat{b}_j$ mode with the appropriate frequency and phase, this nonlinear Hamiltonian (within mean-field and rotating-wave approximations) yields the Hamiltonian in Eq.~(\ref{eq:Hreal2}), with $\Delta = g_0  | \langle \hat{b}_j \rangle|$.  Note that pure three-wave mixing elements can be realized in several different ways in superconducting quantum circuits.  Examples includes the Josephson ring modulator \cite{Abdo2013}, and the recently developed SNAIL Josephson device \cite{Frattini2017}.  
	
	\section{Conclusions}
	In this work, we have introduced and analyzed a bosonic version of the well known fermionic Kitaev chain.  Our system does not require strong photon-photon interactions, but instead exploits the presence of  non-trivial phases in a quadratic, particle non-conserving Hamiltonian.  It exhibits a spatially-asymmetric coupling between local quadrature operators, which in turn results in a variety of remarkable features.  This includes phase-dependent chiral propagation, and a striking sensitivity to boundary conditions impacting both the localization and dynamic stability of system eigenmodes.  Further, despite being non-dissipative, our system has a direct connection to non-Hermitian asymmetric hopping models; this allows us to give our model a topological underpinning.
	
	In terms of outlook, the physics we discuss here could be exploited for applications.  As we have discussed, our system can serve as  unique kind of phase-sensitive amplifier, in which orthogonal quadratures are amplified in opposite directions.  It also is a unique tool for preparing entangled multi-mode Gaussian states.
	
	Our work also suggests directions for future basic research.  We have established a surprising connection between fermionic Majorana modes and bosonic quadrature operators: it is only natural to ask whether this can be taken further.  Future work could thus investigate the role played by symmetry and dimensionality in generalized bosonic Kitaev models.  The role of true photon-photon interactions in this setting would also be extremely interesting.  	
	
	\section*{Acknowledgements}
	
	We thank Kero Lau for useful conversations.  We acknowledge support from NSERC's CGSM program (AM), NSERC (TPB), FQRNT (TPB), the Simons Foundation (Award 505450, AC), and the NSF-MRSEC 1420709 (AM,AC).

	\appendix
	
	\section{Real-Valued Hopping}
	\label{app:Trivial}
	In this appendix, we demonstrate that the same model as in Eq.~(\ref{eq:Hreal}) without any hopping phase is trivial. For the case of an infinite lattice, the Heisenberg equations of motion for the momentum space operators are
	\begin{align}
	\left(i \partial_t - \check{M}_k \right)
	\begin{pmatrix}
	\ha_k \\ \ha^{\dagger}_{-k}
	\end{pmatrix}  
	\hspace{0.2 cm}
	\check{M}_k =
	t \cos(k) \check{\sigma}_z + i \Delta \cos(k) \check{\sigma}_x
	\label{eq:Mk2Defn}
	\end{align}
	The mode energies are readily obtained from the eigenvalues of the dynamical matrix $\check{M}_k$ and are given by $E_{k,\pm} = \pm \sqrt{t^2-\Delta^2}\cos(k)$. The condition for stability $t > \Delta$   is thus the same throughout the band.
	
	To understand why this model displays trivial physics, we now consider a finite chain of $N$ sites with open boundaries. The Hamiltonian is
	\begin{align}
	\hH_B = 
	\frac{1}{2}\sum_{j=1}^{N-1}
	\left(t\ha_{j+1}^\dagger\ha_j
	+i\Delta\ha^\dagger_{j+1}\ha_j^\dagger
	+h.c.
	\right).
	\end{align}
	Translational invariance being broken by the boundary, momentum is not conserved and we can no longer diagonalize the Hamiltonian by moving to a basis of plane waves. Instead, the Hamiltonian is diagonal in the basis of standing waves, which are simply linear combination of plane waves
	\begin{equation}
	\hH_{B} =  \sum_{n = 1}^N 
	\left[
	t \cos k_n  \,
	\ha^\dagger_{k_n} \ha_{k_n} 
	+
	i \frac{\Delta}{2} \cos k_n  \left(
	\ha^{\dagger}_{k_n} \ha_{k_n}^\dagger - h.c.  
	\right)
	\right], 
	\end{equation}
	with $\ha_{k_n}$ a standing wave annihilation operator
	\begin{align}
	\ha_{k_n} = \sqrt{\frac{2}{N+1}}\sum_{j=1}^N\sin(k_n j)\ha_j,
	\end{align}
	and $k_n = n\pi/(N+1)$. The finite chain with open boundary conditions is essentially no different from the infinite lattice or the chain with periodic boundary conditions. One readily sees that stability is achieved throughout the band if and only if $t > \Delta$, just like in the system without boundaries. Furthermore, the eigenstates of the finite sized system are standing waves and thus do no exhibit any localization, unlike the model presented in the main text (c.f. Eqs.(\ref{eq:EigenParticle})-(\ref{eq:EigenHole})).
	
	\section{Arbitrary Hopping Phase}
	\label{app:ArbPhase}
		
	Here, we study a similar Hamiltonian to that in Eq.~(\ref{eq:Hreal}), except now the hopping matrix element has an arbitrary phase $it \to e^{i\phi}t$. As in the main text, we will assume that $t > \Delta$. We find that, depending on the magnitude of the real and imaginary part of the hopping, this more general model exhibits similar physics either to the one presented in the main text, or a trivial tight-binding model. We explain further in what follows.
	
	First, let's assume that $t |\cos(\phi)| < \Delta$. One can easily verify that 
	\begin{align}
	\hH_B & = \frac{1}{2} \sum_{j }
	\left( e^{i\phi } t  \ha^{\dagger}_{j+1} \ha_j + 
	i \Delta  \ha^{\dagger}_{j+1} \ha_{j}^\dagger  + h.c. \right) 
	\nonumber \\
	& = 
	\frac{1}{2}\sum_{j} \left( i t'\hat{\beta}^\dagger_{j+1}\hat{\beta}_j +i\Delta'\hat{\beta}^\dagger_{j+1} \hat{\beta}^\dagger_{j}+h.c.\right),
	\end{align}
	where $t' = t\sin(\phi)$ and $\Delta' = \sqrt{\Delta^2-t^2\cos^2(\phi)}$. The $\hat{\beta}_j$ are Bogoliubov modes of the original photonic operators 
	\begin{align}
	\hat{\beta}_j = \cosh(\xi)\ha_j+i\sinh(\xi)\ha^\dagger_j
	\end{align}
	and the squeeze parameter $\xi$ is defined via 
	\begin{align}
	\tanh(2 \xi) = \frac{t \cos(\phi)}{\Delta}.
	\end{align}
	This is exactly equivalent to the model presented in the main text with renormalized parameters $t \to t \sin(\phi)$ and $\Delta \to \sqrt{\Delta^2-t^2\cos^2(\phi)}$. 
	
	On the other hand, if $t|\cos(\phi)| > \Delta$, the previous squeezing transformation is not well defined. We can no longer map our Hamiltonian to a similar model with purely imaginary hopping phase and renormalized parameters. What we can do instead however, is make a similar transformation that maps the Hamiltonian onto a simple tight binding model. More concretely, we have
	\begin{align}
	\hH_B & = 
	\frac{1}{2} \sum_{j }
	\left( e^{i\phi } t  \ha^{\dagger}_{j+1} \ha_j + 
	i \Delta  \ha^{\dagger}_{j+1} \ha_{j}^\dagger + h.c. \right)
	\nonumber \\
	& = 
	\frac{1}{2}\sum_{j} \left( e^{i \tilde{\phi} }\tilde{t} \hat{\alpha}^\dagger_{j+1}\hat{\alpha}_j+h.c.\right)
	\end{align}
	where, as in the main text, $\tilde{t} = \sqrt{t^2-\Delta^2}$. Here we've introduced the operators $\hat{\alpha}_j$, which are also Bogoliubov modes of the orginal photonic operators 
	\begin{align}
	\label{eq:UniformSqz}
	\hat{\alpha}_j = \cosh(\rho)\ha_j+i\sinh(\rho)\ha_j^\dagger
	\end{align}
	with $\rho$ the squeeze parameter defined via
	\begin{align}
	\tanh(2\rho) = \frac{\Delta}{t\cos(\phi)},
	\end{align}
	and $\tilde{\phi}$ is the phase 
	\begin{align}
	e^{i\tilde{\phi}} = \frac{it\sin(\phi)+\text{sgn}(\cos(\phi))\sqrt{t^2\cos^2(\phi)-\Delta^2}}{\sqrt{t^2-\Delta^2}} .
	\end{align}
	To see why this model does not display any interesting physics, we note that the transformation defined in Eq.~(\ref{eq:UniformSqz}) is valid $\it{regardless}$ of boundary conditions. Thus, in the regime $t|\cos(\phi)| > \Delta$, one can always map the system onto a simple particle conserving tight-binding model. This is in constrast with the model considered in the main text which was unitarily equivalent to an excitation conserving Hamiltonian $\it{only}$ in the case of open boundary conditions. Furthermore, the unitary operator was a local $\it{position}$ $\it{dependent}$ squeezing transformation (c.f. Eq.~(\ref{eq:SqzTrans})), whereas here the relevant operator is spatially uniform.

	\section{Scattering matrix of a regular tight-binding chain}
	\label{app:TBChain}
	For completeness, here we derive the expression of the photon scattering matrix of a regular tight-binding chain $\tilde{s}_{jj'}[\omega]$, from which we immediately obtain the quadrature scattering matrices, c.f. Eqs~(\ref{eq:Sxx_Matrix})-(\ref{eq:Spp_Matrix}). The first step is to compute the Green's function $\mathbf{\tilde{G}}_0[\omega]$ of the unperturbed system, i.e. without the spatially dependent loss. By definition,
	\begin{align}
	\label{eq:DefGreen}
	\mathbf{\tilde{G}}_0[\omega] = \left((\omega+i\frac{\kappa}{2})\mathbf{1}-\mathbf{H}\right)^{-1}
	\end{align}
	where $\kappa$ is the uniform on-site decay rate and $\mathbf{H}$ is the single particle Hamiltonian which, in real space, has matrix elements (c.f. Eq.~(\ref{eq:RegTB}))
	\begin{align}
	\label{eq:HRegTB}
	H_{ij} = i\frac{\tilde{t}}{2}\delta_{i,j+1}-i\frac{\tilde{t}}{2}\delta_{i,j-1}.
	\end{align}
	Note that we could make a local gauge transformation to make the Hamiltonian matrix $\mathbf{H}$ real valued. However, this transformation would also alter the scattering matrix, and we therefore choose to keep the imaginary phase factors in the definition of the Hamiltonian. With Eq.~(\ref{eq:DefGreen}) in combination with Eq.~(\ref{eq:HRegTB}), one easily verifies that the Green's function for a finite chain of $N$ sites is
	\begin{widetext}
		\[
		\tilde{G}_0[j,j';\omega] =  i^{j-j'}\frac{2\sin\left( q[\omega]\min(j,j') \right) \sin \left(q[\omega](N+1-\max(j,j')) \right)}{\tilde{t}\sin \left(q[\omega]\right) \sin \left(q[\omega](N+1)\right)}
		\]
	\end{widetext}
	where $q[\omega]$ is the complex wavevector satisfying the dispersion
	\begin{align}
	\omega+i\frac{\kappa}{2}-\tilde{t}\cos(q[\omega]) = 0.
	\end{align}
	The introduction of spatially dependent loss $\kappa_j$ introduces an effective imaginary potential $\mathbf{V}$ at each lattice site. In real space it has matrix elements
	\begin{align}
	V_{ij} = -i\frac{\kappa_j}{2} \delta_{ij}.
	\end{align}
	The full Green's function $\mathbf{\tilde{G}}[\omega]$ is given by Dyson's equation
	\begin{align}
	\mathbf{\tilde{G}}[\omega] = \mathbf{\tilde{G}}_0[\omega]+\mathbf{\tilde{G}}_0[\omega]\mathbf{V}\mathbf{\tilde{G}}[\omega]  = \frac{1}{\mathbf{1}-\mathbf{\tilde{G}}_0[\omega]\mathbf{V}}\mathbf{\tilde{G}}_0[\omega].
	\end{align}
	Standard input-output theory \cite{Gardiner00,ClerkRMP} then gives a simple relation between the scattering matrix and the Green's function
	\begin{align}
	s_{jj'}[\omega] = \delta_{jj'}-i\sqrt{\kappa_j \kappa_{j'}}\tilde{G}[j,j';\omega]
	\end{align}
	\section{Spatially Varying Hopping and Pairing}
	\label{app:Generalized Model}	
	We now consider a generalized version of our model where the hopping and parametric drive amplitudes vary from bond to bond:
	\begin{align}
	\hH_{B^\prime} \equiv \frac{1}{2}
	\sum_{j=1}^{N-1}  \left(
	-(t_j-\Delta_j) \hx_{j+1} \hp_j + (t_j+\Delta_j) \hp_{j+1} \hx_j 
	\right)
	\end{align}
	 
	As long as $t_j > \Delta_j$ for all $j$, the model can still be mapped to a particle-conserving model. Defining
	\begin{align}
	e^{2r_j} = 
	\frac{t_j+\Delta_j}{t_j-\Delta_j},
	\,\,\,\,
	R_j = \sum_{m=0}^{j-1}r_m,
	\end{align}
	with $r_0$ an arbitrary real number, we make a local position-dependent squeezing transformation: 
	\begin{align}
	\hat{U}\hx_j\hat{U}^\dagger = e^{R_j}\hat{\tilde{x}}_j, 
	\hspace{1 cm}
	\hat{U}\hp_j\hat{U}^\dagger = e^{-R_j}\hat{\tilde{p}}_j
	\label{eq:RjDefinition}
	\end{align}   
	
	One finds that
	\begin{align}
	\hat{U}\hH_{B^\prime}\hat{U}^\dagger 
	& = 
	\frac{1}{2} \sum_{j }
	\left( 
	i \tilde{t}_j  \hta^{\dagger}_{j+1} \hta_j + h.c. \right),
	\label{eq:NonUniformTB}
	\end{align}
	with $\tilde{t}_j = \sqrt{t_j^2-\Delta_j^2}$.
	
	Thus, in the case where $t_j>\Delta_j$ for all $j$, 
	$\hH_{B^\prime}$ is unitarily equivalent to a particle conserving tight-binding chain with a spatially varying tunnel matrix element $i\tilde{t}_j$. This mapping also implies a simple form for the scattering matrices that corresponds to Fig.~\ref{fig:ScatteringEquiv}:
	\begin{align}
	\label{eq:Sxx_Matrix_Disorder}
	s_{jj'}^{xx}[\omega]  = e^{R_j-R_{j'}} \ts_{jj'}[\omega],
	\\
	\label{eq:Spp_Matrix_Disorder}
	s_{jj'}^{pp'}[\omega] = e^{-(R_j-R_{j'})} \ts_{jj'}[\omega]
	\end{align}
	where now $\ts_{jj'}[\omega]$ is the scattering matrix of an $N$-site tight binding chain with hopping matrix elements $i\tilde{t}_j$ and on-site decay rates $\kappa_j$. 

	
	\bibliography{BosMajoranaBib}

\end{document}